# A 30-item Test for Assessing Chinese Character Amnesia in Child Handwriters

Zebo Xu[1], Steven Langsford[1], Zhuang Qiu[1,2]*, Zhenguang Cai[1,3]*

1 Department of Linguistics and Modern Languages, The Chinese University of Hong Kong, Hong Kong SAR

2 Faculty of Humanities and Social Sciences, City University of Macau, Macau SAR

3 Brain and Mind Institute, The Chinese University of Hong Kong, Hong Kong SAR

**Funding**

The research was supported by a GRF grant from the Research Grants Council of the Hong Kong SAR (Grant No. 14613722 to Zhenguang Cai).

**\* Corresponding authors**

Zhuang Qiu

Faculty of Humanities and Social Sciences, City University of Macau, Macau SAR.

zhuangqiu@cityu.edu.mo

Zhenguang Cai

Brain and Mind Institute, Department of Linguistics and Modern Languages, The Chinese University of Hong Kong Shatin, N.T., Hong Kong SAR.

zhenguangcai@cuhk.edu.hk

**Abstract**

Handwriting literacy is an important skill for learning and communication in school-age children. In the digital age, handwriting has been largely replaced by typing, leading to a decline in handwriting proficiency, particularly in non-alphabetic writing systems. Among children learning Chinese, a growing number have reported experiencing character amnesia: difficulty in correctly handwriting a character despite being able to recognize it. Given that there is currently no standardized diagnostic tool for assessing character amnesia in children, we developed an assessment to measure Chinese character amnesia in Mandarin-speaking school-age population. We utilised a large-scale handwriting dataset in which 40 children handwrote 800 characters from dictation prompts. Character amnesia and correct handwriting responses were analysed using a two-parameter Item Response Theory model. Four item-selection schemes were compared: random baseline, maximum discrimination, diverse difficulty, and an upper-and-lower-thirds discrimination score. Candidate item subsets were evaluated using out-of-sample prediction. Among these selection schemes, the upper-and-lower-thirds discrimination procedure yields a compact 30-item test that preserves individual-difference structure and generalizes to unseen test-takers (cross-validated mean $r = .74$ with full 800-item-test; within-sample $r = .93$). This short-form test provides a reliable and efficient tool of assessing Chinese character amnesia in children and can be used to identify early handwriting and orthographic learning difficulties, contributing to the early detection of developmental dysgraphia and related literacy challenges.

**Introduction**

Handwriting is an important skill for both learning and communication (Jones & Hall, 2013). For children, handwriting remains the dominant modality in school learning compared to digital typing (Cutler & Graham, 2008; Jones & Hall, 2013; Sheffield, 1996). Research shows that children acquire letter knowledge more effectively through handwriting than through passive observation (James & Engelhardt, 2012), especially when learning explicit knowledge of visually complex orthographies such as Chinese (Tong & McBride-Chang, 2010), a writing system is also well-known by the lack of mapping between phonology and orthography. Important as it is, Chinese handwriting literacy however has been reported to decline in the digital era due to the lack of handwriting practice (e.g., Xu et al., 2025a; Wang et al., 2020). This shift has given rise to a phenomenon known as character amnesia (referred to as "提笔忘字" in Chinese), where individuals know which character to write but fail to retrieve its full orthography. While existing research has developed tools to assess handwriting ability (Langsford et al., 2024; Xu et al., 2025a; Chow et al., 2003; Tseng & Hsueh, 1997), there are no assessments for evaluating character amnesia rates in Chinese school age population. To fill this gap, we explored a large-scale database of Chinese character handwriting (Xu et al., 2025c) and developed a 30-item test of Chinese character amnesia in children, which can also be used as a reliable and efficient tool to identify early handwriting and orthographic learning difficulties.

**Character amnesia in Chinese handwriting**

The Chinese writing system is renowned for its intricate visual complexity: each character is composed of one or more radicals, which are meaningful subcomponents made up of individual strokes (e.g., the radical 氵 consists of 丶, 丶, and ㇀) and spatially arranged within a square configuration (e.g., 清). Chinese characters exhibit various structural layouts, such as left-right (e.g., 氵 and 青 forming 清) or top-bottom compositions (e.g., 龶 and 月 forming 青). Unlike alphabetic writing system, Chinese has little correspondence between phonology and orthography, making phonological cues unreliable for retrieving written forms. For instance, the character 清 is pronounced *qing₁* (or /tɕʰiŋ/ in IPA), yet its phonemes (/tɕʰ/, /i/, and /ŋ/) do not correspond to any specific graphemes. Instead, each Chinese syllable corresponds to a character that can function as a standalone word (e.g., 清, meaning "clear"). Although some phonetic radicals provide pronunciation hints (e.g., 青 in 清), only about 30% of characters match their phonetic radicals' pronunciations (e.g., Zhu, 2003). As a result, the acquisition of explicit orthographic-motor knowledge for thousands of Chinese characters requires decades of handwriting practice and must be constantly consolidated to prevent attrition (Tong & McBride-Chang, 2010).

Digital typing resembles handwriting in alphabetic scripts, where phonemes are sequentially spelled into graphemes; however, this correspondence does not apply to logographic systems such as Chinese. In Chinese, the dominance of keyboard-based input methods, together with the visual complexity and absence of direct mapping between phonology and orthography, has contributed to the growing prevalence of character amnesia, a phenomenon in which individuals are unable to handwrite a character even though they know what character they are supposed to write (Huang, Zhou et al., 2021; Almog, 2019; Hilburger, 2016). For instance, when asked to handwrite the character from a dictation prompt (e.g., *xia₄qi₂ de qi₂*, 下棋的棋, meaning "chess, as in the phrase *playing chess*"), individuals may correctly access the phonology (i.e., *qi₂*) and meaning (i.e., chess) but fail to produce its orthographic form, only producing a few strokes or none at all. Character amnesia has been widely reported among Chinese speakers (e.g., Lan, 2013; Zhou, 2013; Huang, Lin et al., 2021). In a survey of 2517 participants, 98.8% reported experiencing character amnesia, with 29.5% indicating it occurred regularly (Zhou & Xu, 2013). Similarly, 90% of individuals encountered character amnesia when writing common words (Lan, 2013). More recent empirical analyses of large-scale handwriting datasets reveal that college students exhibit character amnesia in approximately 5.6% of the cases on average (Huang, Zhou, et al., 2021; Wang et al., 2020). These findings suggest that neglecting handwriting practice can contribute to the decay of orthographic representations; however, little is known about handwriting decline among children in the digital era. Our 30-item test provides a rapid (about 10 minutes) and accurate tool for measuring children's character amnesia rates. It enables researchers to examine how handwriting proficiency is influenced by changes in reading and writing habits under the large-scale settings.

**Assessing handwriting in children**

In alphabetic languages, researchers have developed assessment tools to measure children's handwriting literacy. These assessments involve asking children to copy English words and the penmanship (i.e., handwriting legibly and aesthetically) of the copied English words are evaluated by trained testers (Phelps et al., 1985; Reisman, 1999). The number of correctly copied Chinese characters is commonly used to measure handwriting fluency (e.g., Chow et al., 2003; Tseng & Hsueh, 1997). More recent approaches use computerized handwriting input to quantify letter size, spacing, and fluency (e.g., Falk et al., 2010). Adak et al. (2017) further applied convolutional neural networks (CNNs) to handwritten images and human ratings to develop a human-like automatic penmanship assessment.

A range of assessments have been developed to measure Chinese handwriting. For instance, Tseng's Handwriting Speed Test requires children to copy Chinese characters as quickly and accurately as possible, with handwriting fluency evaluated by the number of correctly written characters per minute (Chow et al., 2003; Tseng & Hsueh, 1997). Another assessment developed for kindergarten children, the name-writing test, evaluates handwriting quality on a 9-point scale based on the overall appearance of the written name (Tse et al., 2019). However, existing handwriting assessments in both alphabetic and Chinese writing systems have mainly focused on copying accuracy, speed, or visual quality. They do not capture a handwriter's ability to access and retrieve the orthographic components of characters, a process directly involves character amnesia. To address this limitation, recent studies have employed handwriting-to-dictation tasks, this approach provides a more sensitive measure of orthographic retrieval and has proven useful for assessing character amnesia (Xu et al., 2025a; Xu et al., 2025b; Xu et al., 2025c; Xu et al., 2025d; Langsford et al., 2024; Huang, Lin et al., 2021).

**Assessing character amnesia using item response theory**

Another key challenge in designing such assessments is character selection. Because Chinese literacy curricula require students to learn hundreds of characters, exhaustive testing is impractical. Effective assessment therefore depends on selecting a small yet representative item set. To meet this challenge, Langsford et al. (2024) developed a concise paper-and-pencil assessment of character amnesia. Drawing on a large-scale handwriting dataset in which 42 Mandarin-speaking adults each handwrote 1200 characters (Xu et al., 2025c), they applied IRT (Bock, 1997) to identify 30 characters that were most discriminative of a test-taker's character amnesia rate. By modelling item difficulty and discrimination parameters, IRT enabled the selection of characters that best differentiated between individuals with varying levels of orthographic retrieval ability. Correlational analyses showed that participants' amnesia rates for the 30 selected characters were highly correlated ($r = .89$) with amnesia rates across all 1200 characters, indicating that this test set provides a reliable and efficient estimate of a test-taker's overall character amnesia rate in Chinese handwriting. This finding shows that character amnesia can be measured in short tests with high reliability.

**The present study**

As reviewed above, previous assessments of handwriting literacy have mainly focused on copying fluency (e.g., Skar et al., 2022; Li-Tsang et al., 2022) and penmanship (Xu et al., 2025d). To better evaluate character amnesia, Langsford et al. (2024) applied IRT to a large-scale handwriting database and identified 30 discriminative characters that reliably predicted character amnesia rates across 1200 characters ($r = .89$) among adult handwriters. However, no assessments are available to evaluate character amnesia in child handwriters. This study addresses this gap by applying an IRT model to an existing large-scale handwriting database in which 40 children each handwrote 800 characters in a handwriting-to-dictation task. Using data from a database of handwriting-to-dictation by children (Xu et al., in prep), we developed a short-form test capable of reliably and efficiently distinguishing children's character amnesia rates.

**Child handwriting database**

The data used in this study were borrowed from an existing large-scale handwriting database of child handwriters (Xu et al., in prep). Forty children (23 females, mean age 10.28 years, age range 10-11) in the first semester of the fifth grade at a primary school were recruited to perform a handwriting-to-dictation task. All participants were native speakers of Mandarin and users of simplified Chinese characters. They were right-handed and had normal hearing and normal or corrected-to-normal vision. None of the participants reported any neurological or psychiatric disorders or were undergoing medical treatment.

In this study, 800 stimuli (i.e., 800 dictated characters) were selected from an adult large-scale handwriting database (Xu et al., 2025c). Several filters were applied for stimuli selection: characters with a log frequency between 1.5 and 5.0 (corresponding to 32-100000 corpus occurrences) and more than four strokes were selected to avoid ceiling effects. These filters resulted in a candidate set of 2095 characters. For each character, we extracted the most common bi-character word from SUBTLEX-CH (e.g., 下棋, *xia$_4$qi$_2$*, for target character 棋, *qi$_2$*). These words were then rated for familiarity by 15 participants on a scale from 1 to 7 (the highest value representing the most familiar). These context words were included only when the familiarity scores were equal to or higher than 4 on average, resulting in 1600 candidate characters. These characters were then compiled into dictation phrases (e.g., "下棋的棋", *xia$_4$qi$_2$ de qi$_2$*, the target character "棋", meaning "chess", in the word "下棋", meaning "playing chess"). The audio recordings of the dictation phrases were then generated using Google Text-to-Speech (https://cloud.google.com/text-to-speech). Given that the child handwriters were in the first semester of fifth grade, only characters that were learned before fifth grade were included, based on the primary Chinese textbooks used in compulsory education under the 2011 national curriculum (Cai et al., 2022). These filters resulted in a candidate set of 1035 characters. To reduce the total time for the task, we randomly selected 800 items from this candidate pool. The mean log frequency with regard to the original corpus was 3.55. The number of strokes per character ranged between 5 and 20, with a mean of 9.45, while the number of constituent radicals ranged between 1 and 7, with a mean of 2.91.

The experiment was conducted using the OpenHandWrite_Toolbox (Xu et al., 2025c) on a laptop computer. It comprises two programs: GetWrite and MarkWrite. GetWrite, a WinTab interface integrated within the Psychopy environment (version 1.83.02), supports Wacom Intuos tablets (Wacom PTH-651, Japan). During the experiment, children were seated in front of the laptop and used an inking digitizer pen (Wacom KP-130-00DB, Japan) to write on 8 × 5 grid paper sheets affixed to the Wacom Intuos tablets. Each child participated in four sessions, writing 200 characters per session, resulting in a total of 800 characters per child.

In this experiment, participants first heard a cue sound signalling the start of the experimental trial, followed by a 500 ms blank interval. They then heard a dictation prompt indicating the to-be-written character (e.g., "下棋的棋", *xia$_4$qi$_2$ de qi$_2$*, meaning the target character "棋", chess, in the word "下棋", play chess). Once they finished writing, participants were shown the target character and self-reported whether they had written the character correctly (a correct handwriting response), knew which character was supposed to be written but forgot how to write it (a character amnesia response), or did not understand the dictation prompt (a don't know response). This was followed by another 500 ms blank interval, allowing children to prepare for the next trial (see **Figure 1**).

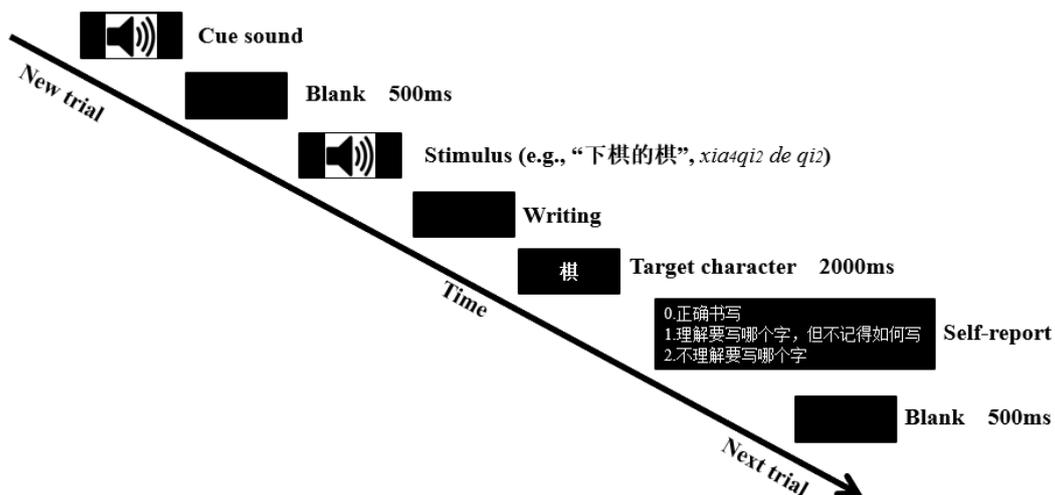

**Figure 1.** The experimental schema was adapted from Xu et al. (2025c). It began with a cue sound "Ding", followed by a 500ms blank interval. Participants then heard the dictation prompt and could immediately begin writing. After finishing handwriting, it was followed by a 2000ms disclosure of the character be written. Participants provided self-report responses by pressing the numbers 0 (correct handwriting), 1 (character amnesia), or 2 (don't know) on the keyboard. Another 500ms blank interval followed before the next trial.

To ensure that children could understand the difference between don't know and character amnesia responses, we explicitly provided the following instructions: don't know response represents they fail to handwrite a target character due to not understanding what character they were supposed to handwrite; for instance, *lian$_2$hong$_2$ de lian$_2$*, which is a plausible pronunciation in Chinese but does not correspond to any meaningful word. In addition, children were told that if they handwrote an incorrect target character by mistake after being shown the target character (e.g., writing the homophone 奇 instead of the target character 棋 in 下棋的棋 this response was also classified as a don't know response rather than a character amnesia response. A character amnesia response indicated that the child knew the target character but could not handwrite it. Providing these specific instructions with concrete examples helped children distinguish character amnesia from don't know responses. Note that there is no objective way for researchers to distinguish between character amnesia and don't know responses; therefore, we relied on participants' self-reports to identify character amnesia responses (which were manually checked; see accuracy checking below).

For accuracy checking, two helpers each checked 400 penscripts (handwritten images) against participants' self-reports using the online survey platform Qualtrics (https://www.qualtrics.com/). For each target character, helpers saw the typed character followed by 40 corresponding penscripts. They classified each as "correct handwriting" if it matched the target character, or "incorrect handwriting" otherwise. A cutoff score for excluding participants was set at 2 standard deviations above the mean questionable self-reports, where they self-reported the penscripts as correct, but helpers identified as incorrect handwriting; or self-reported as incorrect or character amnesia, but helpers identified as a correct response. This process identified three participants with very high ratios of questionable self-reports (26.13%, 32.25%, and 46.38%) and were excluded from the IRT analysis. Among the remaining data after accuracy checking, there were 25916 correct handwriting responses, 1224 character amnesia responses, and 1584 incorrect responses. Only correct handwriting and character amnesia responses (after manual checking) were included in subsequent analyses. Overall, 94.39% of trials showed consistency between participants' self-reports and the manually checked penscripts, indicating high reliability of the self-report data. In addition, we evaluated inter-rater reliability of the manual coding. Another 12 helpers were recruited to check for inter-rater reliability. They were randomly assigned to three group of four raters, each group coded all the penscripts, such that each penscript were coded by three different helpers. We conducted an intraclass correlation analysis to assess coding inter-rater reliability. The results revealed an intraclass correlation coefficient (ICC) of 0.930, with a 95% confidence interval for ICC population values ranging from 0.929 to 0.931, indicating strong agreement among raters.

The mean amnesia rate per participant was 0.04, with a maximum of 0.13. The number of amnesia responses ranged from 0 to 107 across participants. The mean amnesia rate per character was 0.002, with a strongly right-skewed distribution (median = 0.001; third quantile = 0.003; see **Figure 2**). A total of 360 characters never elicited an amnesia response. The trial-level handwriting data is available on the Open Science Framework(https://osf.io/9a8ke/overview?view_only=ef0fcca3f5404786820f9aed199eb813).

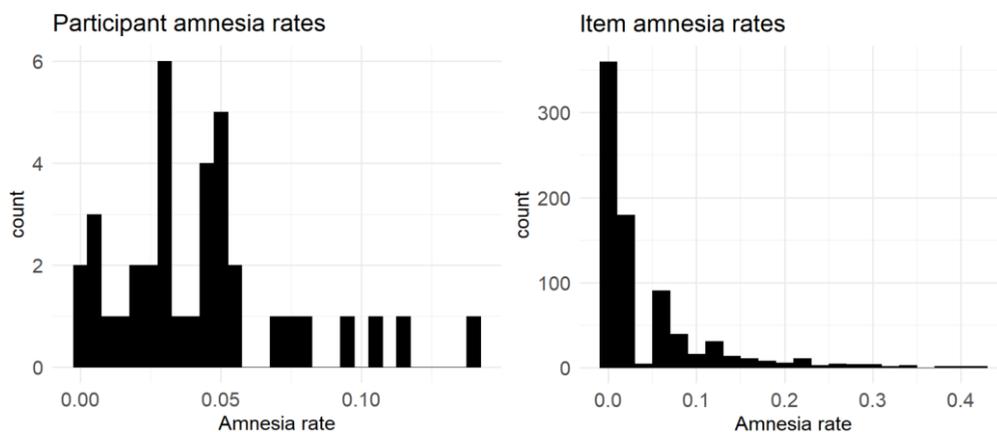

**Figure 2.** Histogram of amnesia rates by-participant and by-item.

**Data analysis**

This session details the item selection procedures, some of which are underpinned by an IRT model. The model adapted a 2-parameter implementation by the Stan development team (2021), estimating participant ability ($\alpha$), item difficulty ($\beta$), and item discriminability ($\gamma$). Each selection procedure has a distinct emphasis: one prioritizes high-discrimination items, another targets a wide spread of difficulty levels, and a third explicitly balances both aims. Random item selection serves as the baseline for comparison. The following three questions guide our analysis:

(1) Given our best estimates of those characteristics, how does prediction performance vary with the number of items in the subset?

(2) How stable are the item-selection procedures given sampling variability in the

estimated item characteristics?

(3) For a 30-item subset, what is the expected out-of-sample performance when predicting character-amnesia rates in future test-takers?

**Model description**

We use a Bernoulli distribution to model each trial, in which the probability of success Pr (success) equals $\text{logit}^{-1}(\gamma_j(\alpha_i-\beta_j))$ for participant i responding to item j. The participant has ability $\alpha$ and the item has difficulty $\beta$ and discrimination $\gamma$. The inverse logit transformation represents the probability of success as an S-shaped curve ranging from zero to one. Participant ability and item characteristics locate the probability of success on the curve, with discrimination $\gamma$ specifying the "slope", while $\alpha$ and $\beta$ specifying the "step". As shown by the subscripts, we estimated distinct participant abilities and item characteristics for each participant and item. We included model implementation details in **Appendix 1**. The final 30-item test is available at https://osf.io/9a8ke/overview?view_only=ef0fcca3f5404786820f9aed199eb813.

**Item selection procedures**

The four item selection procedures described below exclude items with very low variation in responses, which are clearly uninformative. We exclude items with amnesia rates less than 0.01 as too easy, and items that were never written correctly as too difficult. This excluded 360 'easy' items. No items were excluded for being too difficult: the most-forgotten character (靴) had an amnesia rate of 0.419. The selection procedures below draw from the 440 remaining items, and attempt to select subsets that will support accurate inference of amnesia rates in the whole set.

**Random selection**

To create a test with n items, *random selection* chooses uniformly from all available items. We present results from this procedure as a performance baseline.

**Maximum discrimination**

To create a test with n items, this selection procedure ranks eligible items by mean estimated gamma γ (see **Appendix 1**) and selects the top n items. *Maximum discrimination* emphasizes items that give high confidence in how sharply they discriminate between participants above or below some threshold ability. High discrimination items are usually highly informative individually, but collections of the highest discrimination items might be suboptimal if the thresholds they discriminate at are tightly clustered or in rarely seen ability ranges.

**Diverse difficulty**

The *diverse difficulty* selection scheme emphasizes item collections that give full coverage over the range of achievable difficulty levels. To create a test with n items, this selection procedure sorts items by mean estimated beta β (see Appendix 1) and groups them into n bins (with the highest-difficult bin left 'short' if the number of items does not divide evenly). Within each difficulty bin, the highest-discrimination item is chosen.

**Upper-and-lower-thirds**

This procedure follows the 'discrimination score' function in the R package 'psychometric' (Fletcher, 2023; package 'psychometric') which implements a strategy described by Allen and Yen (1979). Under this scheme, the highest and lowest ⅓ of participants are identified and the discrimination score for an item is calculated as:

$$D = \frac{U - L}{N}$$

Where U and L are the number of correct responses in the highest and lowest groups, and N is the size of one of those groups. To create a test with n items, this selection scheme ranks items by D and selects the top n items.

**Effect of the number of items in the test**

For each of the item selection procedures, we evaluated how the predictive performance of the test changes as the test-set size increases. In general, the reliability of a test increases as more items are added to it, but the amount of predictive reliability gained is not uniform (Farkas et al., 2024; McNeish, 2018; Streiner, 2003). To examine the impact of test-set size, we took participants' amnesia rates on the full set of 800 tested characters as the observed ground truth and asked to what extent the amnesia rates estimated from subsets of varying sizes could predict those observed in the full set.

We tested a range of subset sizes from 4 to 70 items for each item selection method, increasing in steps of two items, and computed the correlation coefficient between the predicted and observed *ground truth* amnesia rates for each combination of subset size and selection method. Our focus is on the relative improvement in the correlation coefficient for the selected items as the number of items in the test increases. When the Upper-and-Lower-Thirds procedure was used to select items, the correlation between predicted and *ground truth* amnesia rates approached roughly 0.9 once the test-set size exceeded 20 items, with diminishing improvements for additional items (**Figure 3**). The *Maximum Discrimination* and *Diverse Difficulty* schemes never reached this level of performance, but their correlations stabilized at above 0.75 for a test-set of 30 items and gradually approached 0.8 at larger sizes above 60 items. The *Random Selection* procedure generally performed poorly, as indicated by greater variability in the correlation coefficients. These results suggest that a test-set size of 30 items is adequate for predicting the amnesia rates that would have been observed in the full 800-item test.

**Figure 4** 'unpacks' one of the correlations plotted in **Figure 3** to show predicted versus observed amnesia rates for all participants. This particular test used a test size of 30 items and observed a predicted vs observed correlation of 0.93 for the *Upper-and-Lower-Thirds*, 0.75 for *Maximum Discrimination*, 0.76 for *Diverse Difficulty*, and 0.69 for R*andom* item selection. We present it here as a typical representative of results at this test-set size.

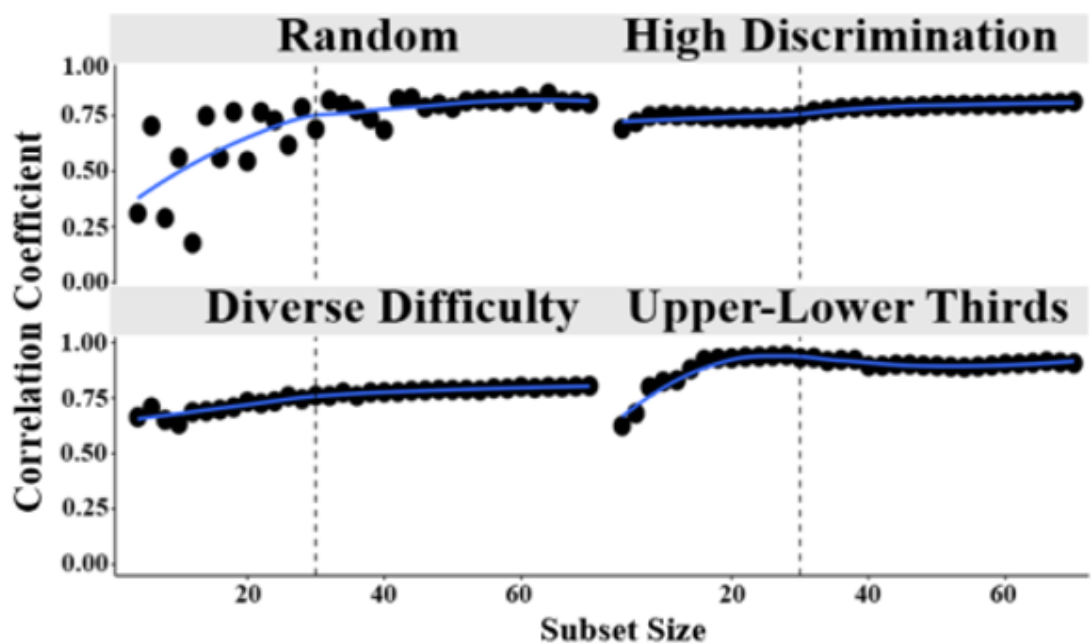

**Figure 3.** Correlations between predicted and observed character amnesia rates using test subsets selected according to the *random*, *maximum discrimination*, *diverse difficulty* and the and *upper-and-lower-thirds* method. A proposed test-set size of 30 items is highlighted with a dashed vertical line.

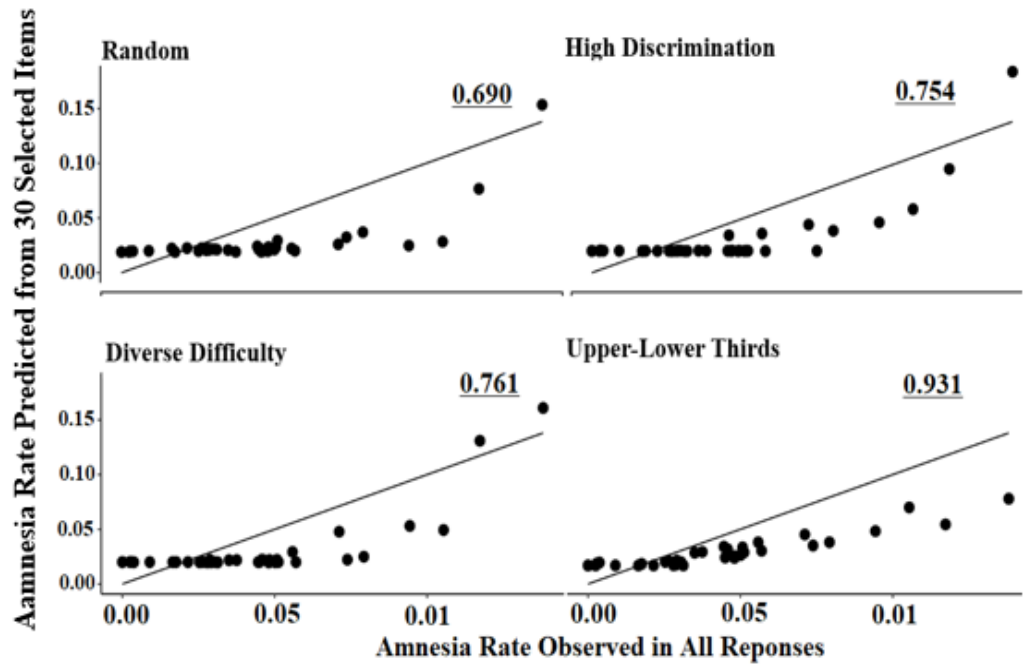

**Figure 4.** A single prediction check at the target test size of 30 items.

**Stability of the item-selection procedures**

To evaluate the stability of the item selection procedures described in the previous sections, we repeatedly removed eight participants and selected a subset of 30 items based on the item characteristics estimated from the data of the remaining 27 participants. Across 30 repetitions of participant removal and item selection using each of the four methods, we assessed the extent to which the selected item subsets remained stable throughout the resampling process.

It is evident that the *Random Selection* procedure is orthogonal to the characteristics of the participants involved, whereas the *Maximum Discrimination*, *Diverse Difficulty*, and Upper-and-Lower-Thirds procedures produce subsets of character items deterministically contingent on the participants included in the calibration data. In this way, the stability of an item's *inclusion rate* across repeated resampling indicates how sensitive the item selection procedure is to sampling variability.

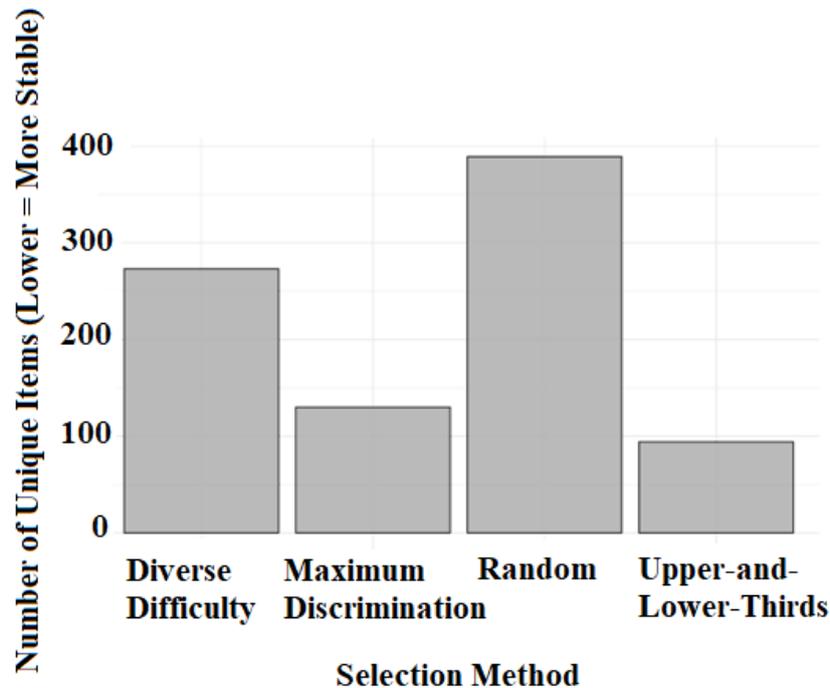

**Figure 5.** Item selection stability across four methods. Bars show the number of distinct items selected across 30 resampling iterations. Lower values indicate greater stability. The Upper-and-Lower-Thirds and *Maximum Discrimination* procedures demonstrate high stability compared to the *Diverse Difficulty* and *Random Selection* methods.

Across the 30 repetitions, the Upper-and-Lower-Thirds item selection procedure used only 94 distinct characters, with 24 characters appearing in more than half of the repetitions. The *Maximum Discrimination* procedure used 130 distinct characters in the 30 subsets it selected, with 27 characters appearing in more than half of the subsets. These substantial overlaps indicate that increasing the number of calibration participants further is unlikely to result in substantial changes to the subsets selected by the Upper-and-Lower-Thirds and *Maximum Discrimination* methods. By comparison, the *Diverse Difficulty* item selection procedure used 273 distinct characters, and none of these characters appeared in more than half of the subsets.

**Performance with new test-takers**

We evaluated the out-of-sample performance of 30-item tests derived from four different item-selection procedures. In each of 30 repetitions, we sampled 10 participants and held out their data. Item characteristics were estimated using the data of the remaining 25 participants, and we selected 30-item subsets according to the four different selection procedures described in the previous section. We then compared the amnesia rates of the 10 held-out participants estimated from the 30-item subsets with their observed ground truth amnesia rates based on the full set of 800 items.

Table 1. Correlation between the observed and predicted amnesia rates for the held-out participants across different item-selection procedures.

| Selection Procedure | N | Mean | SD | Min | Max | SE | CI_lower | CI_upper |
|---|---|---|---|---|---|---|---|---|
| Diverse Difficulty | 30 | 0.35 | 0.32 | -0.12 | 0.87 | 0.06 | 0.23 | 0.47 |
| Maximum Discrimination | 30 | 0.68 | 0.20 | -0.04 | 0.92 | 0.04 | 0.61 | 0.75 |
| Random | 30 | 0.53 | 0.30 | -0.42 | 0.91 | 0.06 | 0.42 | 0.65 |
| Upper-and-Lower-Thirds | 30 | 0.74 | 0.15 | 0.37 | 0.96 | 0.03 | 0.69 | 0.80 |

The correlation between the observed and predicted amnesia rates for the held-out participants is shown in **Table 1** and **Figure 6**. Over thirty repetitions, the correlation coefficients ranged between 0.37 and 0.96 for Upper-and-Lower-Thirds item sets with a mean of 0.74 and a 95% confidence interval (95% CI) between 0.69 and 0.80. The correlation coefficients of the *Maximum Discrimination* item sets ranged between -0.04 and 0.92, with a mean of 0.68 and a 95% CI between 0.61 and 0.75, slightly overlapping with the 95% CI of the Upper-and-Lower-Thirds item sets (**Table 1**). The subsets selected using the *Diverse Difficulty* and Random procedures did not perform well for held-out participants. Their mean correlations were 0.35 and 0.53, respectively, and their standard error roughly doubled that of the Upper-and-Lower-Thirds item sets.

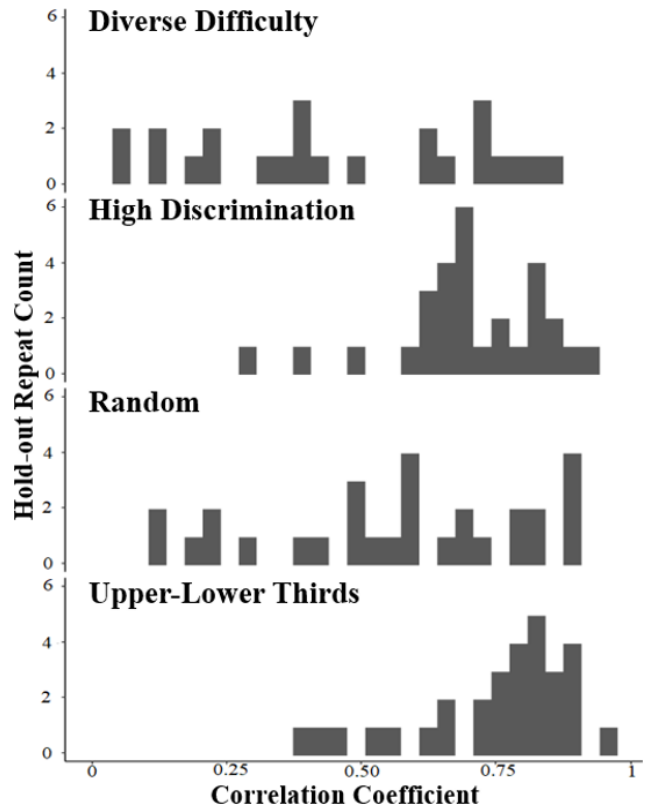

**Figure 6.** Histograms show the 30 correlations obtained between estimates of the 10 hold-out participants' amnesia rates based only on the selected subset of items and the observed amnesia rates taken from their responses to all items.

**General discussion**

This study developed and validated a short-form test to assess Chinese character amnesia in school age children based on a large-scale handwriting dataset, in which 40 children wrote 800 characters from dictation (Xu et al., in prep). Among four item selection procedures (i.e., random selection, maximum discrimination, diverse difficulty, and upper-and-lower-thirds), the Upper-and-Lower-Thirds method yielded the best performance, achieving a correlation of 0.93 between predicted and observed amnesia rates (**Figure 4**). Cross-validation, involving repeated holdouts of 10 participants and refitting the model on the remaining 25, showed high generalizability to unseen test-takers ($r = 0.74$; 95% CI = 0.69 - 0.80). This strategy produced the most stable calibration estimates, indicating that the 30-item test reliably captures individual differences in handwriting literacy. Extending earlier work on adult handwriters (Langsford et al., 2024), the present study provides an early literacy assessment capable of efficiently quantifying orthographic attrition among children.

A key methodological contribution of this work lies in the application of out-of-sample validation to evaluate model generalizability - an approach well established in machine learning (e.g., Xu et al., 2025b) but rarely used in development of assessment using IRT. The application of out-of-sample validation utilised in this study repeatedly calibrated and estimated item characteristics for all 800 characters. Using these estimates, item subsets were selected for each repetition according to each of the three selection procedures. Then the model was re-fit to a new test-taker and finally calculated the correlation between the observed amnesia rates from the full set and the predicted character amnesia rates from the selected subset of items. By repeatedly holding out participants and testing the model's predictive accuracy for unseen individuals, we minimized overfitting and ensured that the 30-item short form can be applied confidently in large-scale settings. This methodological rigor offers stronger empirical support for the predictive validity and real-world applicability of the test.

Another important application of our test is the exploration of demographic and environmental factors contributing to character amnesia among child handwriters. Previous research has shown that individuals are more susceptible to character amnesia when they spend more time using digital devices and less time writing by hand or reading printed materials (Huang et al., 2021). These findings support the view that character amnesia arises from weakened orthographic representations due to insufficient reading and handwriting experience. However, it remains unclear whether character amnesia in children results primarily from reduced handwriting practice or from the direct use of digital input methods. Addressing this question requires large-scale assessments of handwriting literacy. Our 30-item test provides a rapid and scalable means to assess character amnesia in children and can be easily integrated into online surveys for large-scale demographic data collection. This approach allows researchers to examine how factors such as age, gender, pen exposure, print exposure, digital exposure, and input method contribute to character amnesia in children. For instance, phonology-based input methods such as Pinyin allow users to produce characters without engaging orthographic components, potentially undermining the orthographic-motor representations. If so, children who use orthography-based input systems such as Cangjie may show greater resilience against character amnesia and related handwriting difficulties. Findings from such research can provide valuable insights into the predictors of handwriting decline in the digital era, guiding policymakers and educators in developing targeted interventions to preserve Chinese handwriting literacy. Ultimately, this work contributes to understanding and safeguarding an essential component of the Chinese language and cultural heritage.

Our assessment has the potential to pre-screening developmental dysgraphia, identifying children with high rates of character amnesia (Xu et al., 2026; McCloskey & Rapp, 2017; McBride, 2019). Previous research has developed assessment tools for handwriting literacy by using copying task (e.g., Skar et al., 2022; Li-Tsang et al., 2022; Martínez-García et al., 2021). However, copying tasks typically minimize the need for orthographic retrieval, which may fail to detect children with difficulties in retrieving orthographic representations. In contrast, the handwriting-to-dictation task, in which individuals write characters based on spoken phrases, provides a more direct measure of orthographic retrieval ability (Xu et al., 2025a; Xu et al., 2025b; Xu et al., 2025c). The development of this test will also require re-estimation of item characteristics and performance norms for both children with developmental dysgraphia and typically developing peers. Moreover, it should be calibrated for each grade level to accurately distinguish children with dysgraphia from those with typical handwriting development. Such work can inform educational policy and clinical practice, enabling timely and cost-effective support for children at risk of handwriting difficulties.

Our item selection and validation procedures can be adapted to create similar assessment tools for other writing systems, including both logographic (e.g., Japanese kanji, traditional Chinese) and alphabetic scripts (e.g., English). Cross-linguistic application of this approach may reveal whether orthographic attrition across languages shared a similar neurocognitive mechanism across writing systems (a language-general view; Zhao et al., 2017; Nakamura et al., 2012) or whether distinct neural processes are involved (a language-specific view; Murphy et al., 2019; Tan et al., 2005). It may reveal common vulnerabilities in orthographic processing and how they deteriorate in the digital era. Such insights can inform global educational policies for maintaining handwriting proficiency and supporting the preservation of complex orthographic systems. Nevertheless, item selection procedures may differ across populations, educational settings, or socioeconomic contexts. Larger and more diverse samples will be required to develop the short-form assessment into a standardised large-scale screening.

**Conclusion**

Social media reports and empirical studies have highlighted the growing prevalence of character amnesia in Chinese handwriting in the digital era. Existing handwriting assessments for children often rely on copying tasks, which minimize the need to retrieve orthographic representations and thus fail to capture character amnesia rates. To address this issue, we developed a 30-item test that accurately and effectively assesses Chinese character amnesia rate in school age children. Cross-validation using repeated participant holdouts showed that the 30-item test correlated strongly with the full 800-item test. This study provides a foundation for developing computational tools to evaluate early handwriting literacy and support the early detection of handwriting difficulties in educational settings.

**Appendix 1: Model implementation**

We fit a two-parameter (IRT) model in which the probability of a correct response on each dictation trial is modelled as a Bernoulli random variable. The probability of success for each trial depends on participant i and item j, Pr(success) = logit$^{-1}$($\gamma_j$($\alpha_i$ − $\beta$)), where α denotes participant ability, β item difficulty, and γ item discrimination. Ability parameters α were assigned a standard normal prior, with half-Cauchy priors for the scale of γ and β. Discrimination parameters were constrained to be positive via a log-normal prior (see **Figure 7**).

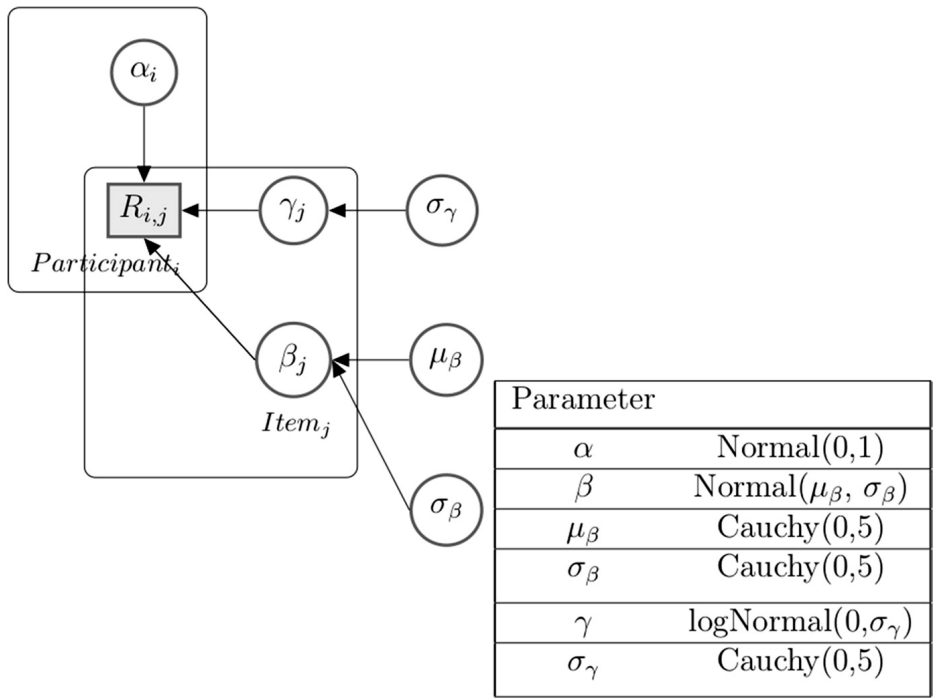

**Figure 7.** 2-parameter IRT model borrowed from Langsford et al. (2024). The model is specified hierarchically, allowing for partial pooling across items and participants. The parameter α represents participant ability, β represents item difficulty and γ represents item discriminability.